\input epsf.sty
\input ash.sty

\documentstyle[referee,psfig]{mn}

\newif\ifAMStwofonts

\newcommand{\m}{$\mu$m}
\newcommand{\elI}{$\epsilon \propto \lambda ^{-1}$}
\newcommand{\elII}{$\epsilon \propto \lambda ^{-2}$}
\newcommand{\ms}{M{\tiny $\odot$}}
\newcommand{\ls}{L{\tiny $\odot$}}

\newcommand{\hii}{H\,{\sc ii}}

\ifoldfss
  \ifCUPmtlplainloaded \else
    \NewTextAlphabet{textbfit} {cmbxti10} {}
    \NewTextAlphabet{textbfss} {cmssbx10} {}
    \NewMathAlphabet{mathbfit} {cmbxti10} {} 
    \NewMathAlphabet{mathbfss} {cmssbx10} {} 
  \fi
  \ifAMStwofonts
    \ifCUPmtlplainloaded \else
      \NewSymbolFont{upmath} {eurm10}
      \NewSymbolFont{AMSa} {msam10}
      \NewMathSymbol{\upi}     {0}{upmath}{19}
      \NewMathSymbol{\umu}     {0}{upmath}{16}
      \NewMathSymbol{\upartial}{0}{upmath}{40}
      \NewMathSymbol{\leqslant}{3}{AMSa}{36}
      \NewMathSymbol{\geqslant}{3}{AMSa}{3E}

      \let\leq=\leqslant 
      \let\geq=\geqslant 
    \fi
  \fi
\fi 

\ifnfssone
  \newmathalphabet{\mathit}
  \addtoversion{normal}{\mathit}{cmr}{m}{it}
  \addtoversion{bold}{\mathit}{cmr}{bx}{it}
  \newmathalphabet{\mathbfit} 
  \addtoversion{normal}{\mathbfit}{cmr}{bx}{it}
  \addtoversion{bold}{\mathbfit}{cmr}{bx}{it}
  \newmathalphabet{\mathbfss} 
  \addtoversion{normal}{\mathbfss}{cmss}{bx}{n}
  \addtoversion{bold}{\mathbfss}{cmss}{bx}{n}
  \ifAMStwofonts
    \ifCUPmtlplainloaded \else
      \UseAMStwoboldmath
      \makeatletter
      \new@mathgroup\upmath@group
      \define@mathgroup\mv@normal\upmath@group{eur}{m}{n}
      \define@mathgroup\mv@bold\upmath@group{eur}{b}{n}
      \edef\UPM{\hexnumber\upmath@group}
      \new@mathgroup\amsa@group
      \define@mathgroup\mv@normal\amsa@group{msa}{m}{n}
      \define@mathgroup\mv@bold\amsa@group{msa}{m}{n}
      \edef\AMSa{\hexnumber\amsa@group}
      \makeatother
      \mathchardef\upi="0\UPM19
      \mathchardef\umu="0\UPM16
      \mathchardef\upartial="0\UPM40
      \mathchardef\leqslant="3\AMSa36
      \mathchardef\geqslant="3\AMSa3E

      \let\leq=\leqslant 
      \let\geq=\geqslant 
    \fi
  \fi
\fi 

\ifnfsstwo
  \DeclareMathAlphabet{\mathbfit}{OT1}{cmr}{bx}{it}
  \SetMathAlphabet\mathbfit{bold}{OT1}{cmr}{bx}{it}
  \DeclareMathAlphabet{\mathbfss}{OT1}{cmss}{bx}{n}
  \SetMathAlphabet\mathbfss{bold}{OT1}{cmss}{bx}{n}
  \ifAMStwofonts
    \ifCUPmtlplainloaded \else
      \DeclareSymbolFont{UPM}{U}{eur}{m}{n}
      \SetSymbolFont{UPM}{bold}{U}{eur}{b}{n}
      \DeclareSymbolFont{AMSa}{U}{msa}{m}{n}
      \DeclareMathSymbol{\upi}{0}{UPM}{"19}
      \DeclareMathSymbol{\umu}{0}{UPM}{"16}
      \DeclareMathSymbol{\upartial}{0}{UPM}{"40}
      \DeclareMathSymbol{\leqslant}{3}{AMSa}{"36}
      \DeclareMathSymbol{\geqslant}{3}{AMSa}{"3E}

      \let\leq=\leqslant 
      \let\geq=\geqslant 
    \fi
  \fi
\fi 

\ifCUPmtlplainloaded \else
  \ifAMStwofonts \else 
    \def\upi{\pi}
    \def\umu{\mu}
    \def\upartial{\partial}
  \fi
\fi
\title[Star formation in RCW 106]
{Study of star formation in RCW 106 using 
far infrared observations} 

\author[A.D. Karnik et al.]
{A.D.~Karnik, S.K.~Ghosh, T.N.~Rengarajan and R.P.~Verma\thanks{
Send all correspondences to : vermarp@tifr.res.in} \\
Tata Institute of Fundamental Research, Homi Bhabha Road, 
Mumbai (Bombay) 400 005, India}

\date{2001 April 4}

\pubyear{2001}

\begin{document}

\maketitle

\label{firstpage}

\begin{abstract}
High resolution far-infrared observations of a large area of the star forming
complex RCW 106 obtained using the TIFR 1m balloon-borne telescope 
are presented. Intensity maps have been  obtained simultaneously in two 
bands centred around 150 and 210 $\mu$m. Intensity maps have also been
obtained at the four {\it IRAS} bands using HIRES processed {\it IRAS} data. 
From the 150 and 210 $\mu$m maps, reliable maps of 
dust temperature and optical depth have been generated. 
The star formation in this complex has  occurred in five 
linear sub-clumps. Using the map at 210 $\mu$m, which has a spatial 
resolution superior to that of the {\it IRAS} at 100 $\mu$m, 23 sources 
have been identified. The spectral energy distribution (SED) and 
luminosity of these sources have been determined using the associations 
with the {\it IRAS} maps. Luminosity distribution of these sources has been
obtained. Assuming these embedded sources to be ZAMS stars and
using the mass-luminosity relation for these, the power law slope of 
the initial mass function is found to be $-1.73\pm0.5$. This index for 
this very young complex is about the same as that for more evolved 
complexes and clusters. Radiation transfer calculations in 
spherically symmetric geometry have been undertaken to fit the 
SEDs of 13 sources with fluxes in both the TIFR and the {\it IRAS} bands. 
From this, the r$^{-2}$ density distribution in the envelopes is ruled out. 
Finally, a correlation is seen between the luminosity of embedded 
sources and the computed dust masses of the envelopes.

\end{abstract}


\section{INTRODUCTION}    

Far-Infrared (FIR) observations provide an important tool to study formation 
of high mass stars, especially at a very early stage of their formation 
when they are deeply embedded in their parent clouds. The TIFR group has
been active in observing in the FIR, especially at trans-{\it IRAS} wavelengths, 
using the TIFR 1m balloon-borne telescope (Ghosh et al. 1988; Ghosh et al.
1989a; Ghosh et al. 1989b; Ghosh et al. 1990; Ghosh et al. 2000;
Verma et al. 1994; Mookerjea et al. 1999; Mookerjea et al. 2000). 
Taking advantage of our location, we have been mapping several star 
forming complexes in the southern hemisphere, not easily accessible from
the higher latitudes of northern hemisphere. 
Balloon-borne telescope is also efficient in mapping a large area with 
moderate sensitivity. In this paper we describe our study of high mass 
star formation in RCW 106, a large southern star forming complex. 
This complex, harbouring a score or more of high mass stars, provides 
us an opportunity to study the mass function of stars at a very early 
stage and the physical conditions of their envelopes. This is  aided by 
our simultaneous mapping at 150 and 210 $\mu$m with a high spatial resolution 
of $\sim 1\arcmin$ enabling us to resolve more sources than the {\it IRAS}
at 100 $\mu$m.

RCW 106 is an optically visible \hii\ region in the southern Galactic plane.
The distance of this \hii\ region is 3.6 kpc \cite{1979ApJ...232..761L}.
It was discovered by Rodgers, Campbell \& Whiteoak (1960)
while they were surveying the southern milky way in H$\alpha$ line emission.
Radio continuum emission towards this region has been mapped by Shaver
\& Goss
(1970a) and Goss \& Shaver (1970) at 408 MHz and 5000 MHz. Interferometric 
observations of a part of the region are available at 1415 MHz (Retallack \& 
Goss, 1980), 1420 MHz (Forster et al. 1987), 6.67 GHz and 8.64 GHz
(Walsh et al. 1998). The associated molecular cloud was detected by 
Gillespie et al. (1977) during 
their observations of southern Galactic \hii\ regions in the 
J = 1--0 transition of CO. Similarly, interstellar molecules like
OH \cite{CaswellHaynes75}, H$_2$O
\cite{Kaufmann.etal.77}, NH$_3$ \cite{Batchelor.etal.77}, CS 
\cite{GardnerWhiteoak78} and H$_2$CO \cite{GardnerWhiteoak84}
were detected towards this region  
indicating extremely dense molecular material. 
Several maser sources in the lines of    
H$_2$O (Batchelor et al. 1980;
Braz \& Scalise 1982;
Braz et al. 1989;
Scalise, Rodriguez \& Mendoza-Torres 1989),
OH \cite{Caswelletal80} and methanol 
(Macleod \& Gaylard 1992; 
Schutte et al. 1993; 
Caswell et al. 1995; 
van der Walt, Gaylard \& Macleod 1995;
van der Walt et al. 1996; 
Ellingsen et al. 1996; 
Macleod et al. 1998) 
have been detected towards this complex  in independent surveys 
and  observations towards  selected {\it IRAS} sources.
All these  suggest ongoing high mass star formation in the region.

In the next section we describe the observations and details of the photometer 
employed. In Sec.\ 3 the intensity maps in the two bands and the derived maps of 
dust temperature and optical depth are presented. Intensity maps from HIRES
processed {\it IRAS} observations are also  presented in the four {\it IRAS} bands.
Sources extracted from these maps and the spectral energy distribution
(SED) derived from our and {\it IRAS} observations  as well as the comparison of
FIR sources with the radio continuum sources
are also presented. Sec.\ 4 presents a discussion of the FIR luminosity 
distribution and the initial mass function (IMF) derived from this. 
Sec.\ 5 describes the physical properties of the envelopes 
of selected sources based on radiation transfer modelling.    
The results are summarised in the last section. More details of this study
can be found in the PhD thesis of Karnik (Karnik 2000).

\section{Observations}    

\begin{figure*}
\centerline{\psfig{angle=0,figure=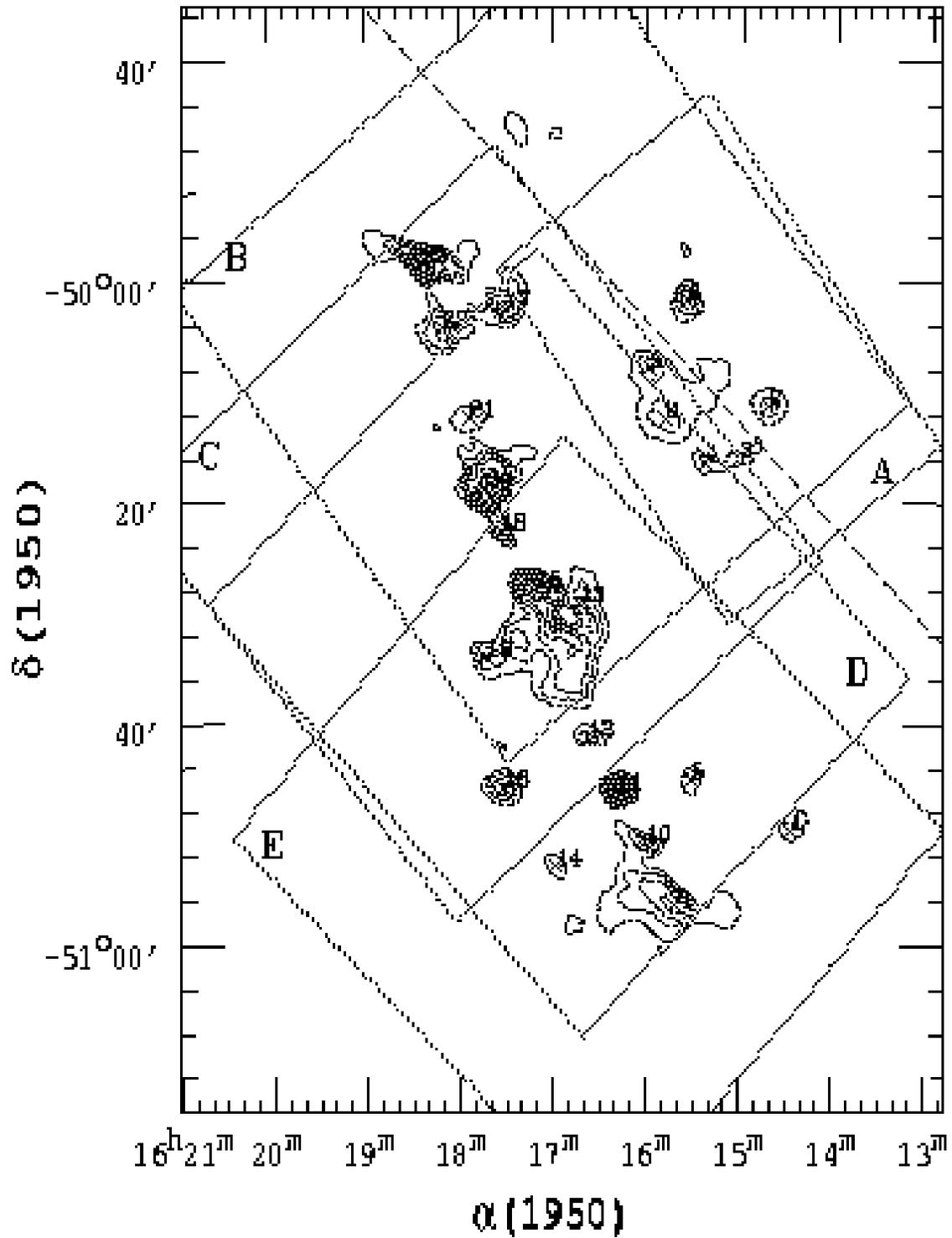,width=16.5cm}}
\caption{ 
The map of RCW 106 region  at 210 \micron .
Contour levels are drawn at 1, 2.5, 5, 10, 20, 30, 50, 70 and 90 
percent of peak intensity  (4375 Jy /\sqam\ ).
The rectangular boxes show the approximate area covered by each of the
individual raster scans A, B, C, D and E respectively. 
The dashed line shows the position of the Galactic plane.
The crosses show  positions of the detected sources.
\label{rcwcoversrc} }
\end{figure*}

\subsection{Balloon-borne FIR observations}

The observations were carried out on February 20, 1994 using a two band  array 
photometer in the focal plane of the TIFR 1m balloon-borne telescope which
 was launched from the TIFR Balloon Facility, Hyderabad, India. Details of 
the telescope and observing procedures have been described by Ghosh et al. 
(1988). 
The photometer used has  been described by Verma, Rengarajan \& Ghosh  
(1993). It consists of 
a pair of compact arrays each having 2 $\times$ 3 silicon bolometers  cooled 
to 0.3 K by liquid $^{3}$He . The field of view of each detector 
was $1\farcm6$. The sky was chopped along the cross elevation axis with a 
throw of $4\farcm2$. Two bands centred around 150 and 210 $\mu$m were 
obtained by splitting the incoming beam using a cooled dichroic filter and 
other band limiting filters. The sky viewed in the two bands was almost 
identical. Jupiter was used for flux calibration as well as for obtaining 
the point spread function (PSF). A large area of about 1.2 $\sq\arcdeg$  of RCW 
106 complex was mapped in 
five overlapping rasters.

All six detectors of each band were independently calibrated for
their responsivity by using in situ measurements of Jupiter.
The telescope aspects corresponding to different detectors of each band
were reduced to a reference detector using the known relative positions in
the focal plane.
 The responsivity corrected signals 
(during the mapping observations of the target
source), from all detectors of the same band were then 
gridded into a 
two dimensional sky matrix ( elevation $\times$ cross elevation; cell
size of $0\farcm3$ $\times$ $ 0\farcm3$).
This signal matrix was 
deconvolved using a procedure based on maximum entropy 
method (MEM; see Ghosh et al. 1988
for details). The FWHMs of the deconvolved beam for Jupiter were found to
be $1\arcmin$ $\times$ 1$\farcm3$ (elevation $\times$ cross elevation) for both 
bands.

Since our wavelength bands are broad, the flux density estimated from the 
observed signal is a function of the incident spectrum. Following the {\it IRAS} 
convention we present flux densities for a $\nu f_{\nu}$ = constant spectrum. 
When computing flux density for other spectral shapes a colour correction 
is applied such that $f_{\nu}$(actual) = $f_{\nu}$(quoted)/ K. For a 
gray body spectrum with emissivity \elI\ the K factor at 150 $\mu$m varies 
from 0.85 to 0.57 as the temperature is varied from 20 K to 60 K; 
at 210 $\mu$m the variation is from 0.84 to 0.85. For an emissivity 
\elII\, the corresponding values are 0.7 to 0.52 for the 150 $\mu$m band and 
0.81 to 0.87 for the 210 $\mu$m band. 

\subsection{ Pointing accuracy}

The absolute pointing accuracy was determined in two different ways. 
(i) An optical photometer (with a photomultiplier tube as detector)
at the cassegrain focal plane of the telescope
always views a region of the sky neighbouring the FIR field
(Naik et al. 2000).
The sky chopped signals obtained from this photometer during the FIR scans 
across the target source were also MEM deconvolved 
to generate optical intensity maps of the scanned region.
The stars detected in these maps have been used to quantify absolute
aspects of our maps. Using 17 well identified and isolated 
optical stars, the mean deviation and rms were found to be 0$\farcm$2 and 
0$\farcm$7 respectively along RA and 0$\farcm$06 and 0$\farcm$68 along the  
Dec axis. 
(ii) A similar exercise was carried out using the coordinates
of six well isolated FIR sources in the {\it IRAS} as well as the TIFR maps.
The rms deviations along RA and Dec axes were found to be 0$\farcm45$
 and 0$\farcm2$ respectively. It may be mentioned that the FIR signal 
matrix has a much higher filling factor as compared to the optical
signal matrix generated using only a single detector.
Thus, our absolute pointing over an area 
as large as 1.2 $\sq\arcdeg$ observed over an hour is better than 
0$\farcm5$. 

\subsection{IRAS-HIRES maps}

The HIRES maps processed from the {\it IRAS} data of RCW 106 region were obtained
 from the Infrared Processing and Analysis Center (IPAC\footnote{IPAC is
funded by NASA as part of the part of the {\it IRAS} extended mission
program under contract to JPL.}, Caltech).
These maps were processed using the maximum correlation 
method (MCM, Aumann, Fowler \& Melnyk 1990). 
The  resolution enhancement results in
 elliptical beams with FWHMs of  $1\farcm$0 $\times$ $0\farcm$45, 
0$\farcm$95 $\times$ 0$\farcm$45, 1$\farcm$3 $\times$ 0$\farcm7$, and  
1$\farcm$9 $\times$ 1$\farcm$3 at 12, 25, 60 and 100 $\mu$m respectively. 
The position angle of the major axis of the beam is 100$\arcdeg$. 
The overall variation in the beam size in different regions of the map
 is about 15\%.

\section{Results}   

\subsection{Intensity maps}

\begin{figure*}
\centerline{\psfig{angle=0,figure=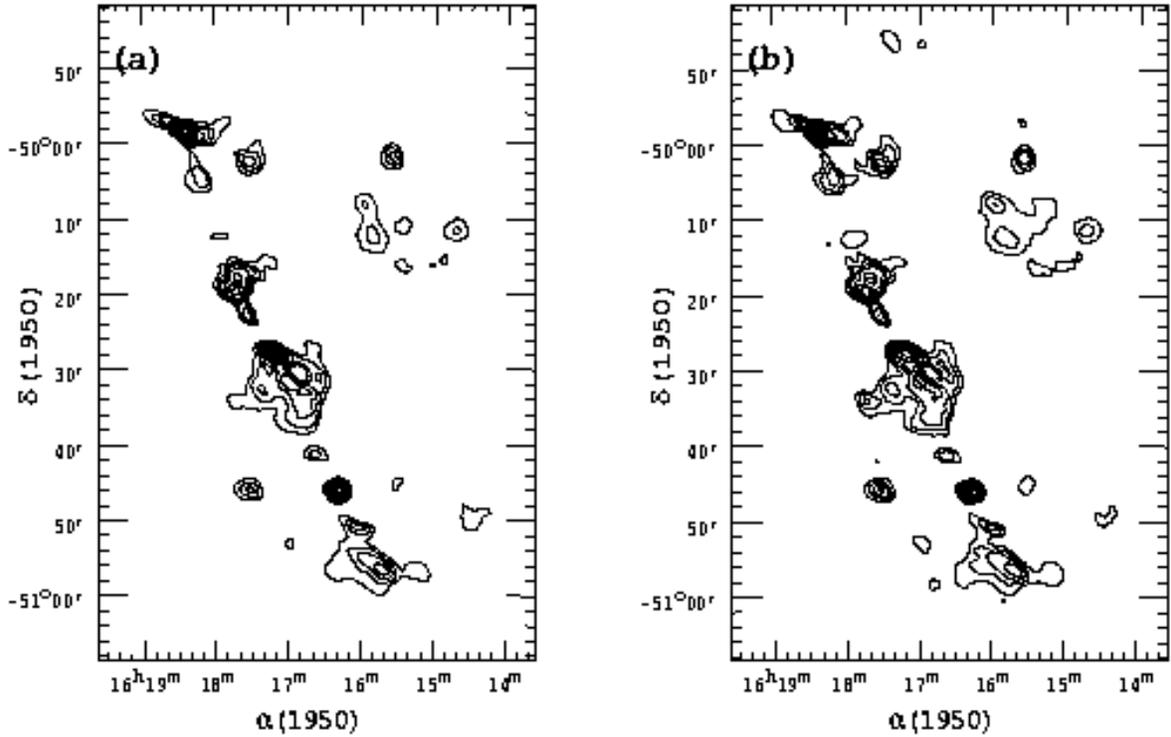,width=16.5cm}}

\caption{ TIFR maps of RCW 106 region for (a) 150 \& (b) 210 \micron.
Contour levels are drawn at 1, 2.5, 5, 10, 20, 30, 50, 70 and 90 
percent of peak intensity  9728 and 4375 Jy /\sqam\ at 
150 and 210 \micron\ respectively.
\label{rcwtifr} }
\end{figure*}

Fig. \ref{rcwcoversrc} shows an enlarged intensity map at  210 $\mu$m 
obtained from our observations combining all the five rasters. The 
boundaries of individual raster areas are also shown. The long dashed 
line shows the position of the Galactic plane. In Fig. \ref{rcwtifr}a 
and \ref{rcwtifr}b we show the intensity maps at 150 $\mu$m and 210 $\mu$m.  
The peak flux densities at 150 and 210 $\mu$m are 9728 and
 4375 Jy/$\sq\arcdeg$ respectively while the noise is found to be 16 and
 9 Jy/$\sq\arcdeg$. The lowest contour level  shown is 1\% of the peak 
and is several times the noise. From the figures it is seen that the 
star formation in RCW 106 is along a narrow ridge roughly parallel to 
the Galactic plane. It is also seen that the FIR emission is subdivided
 into  five  collinear sub complexes.

\begin{figure*}
\centerline{\psfig{angle=0,figure=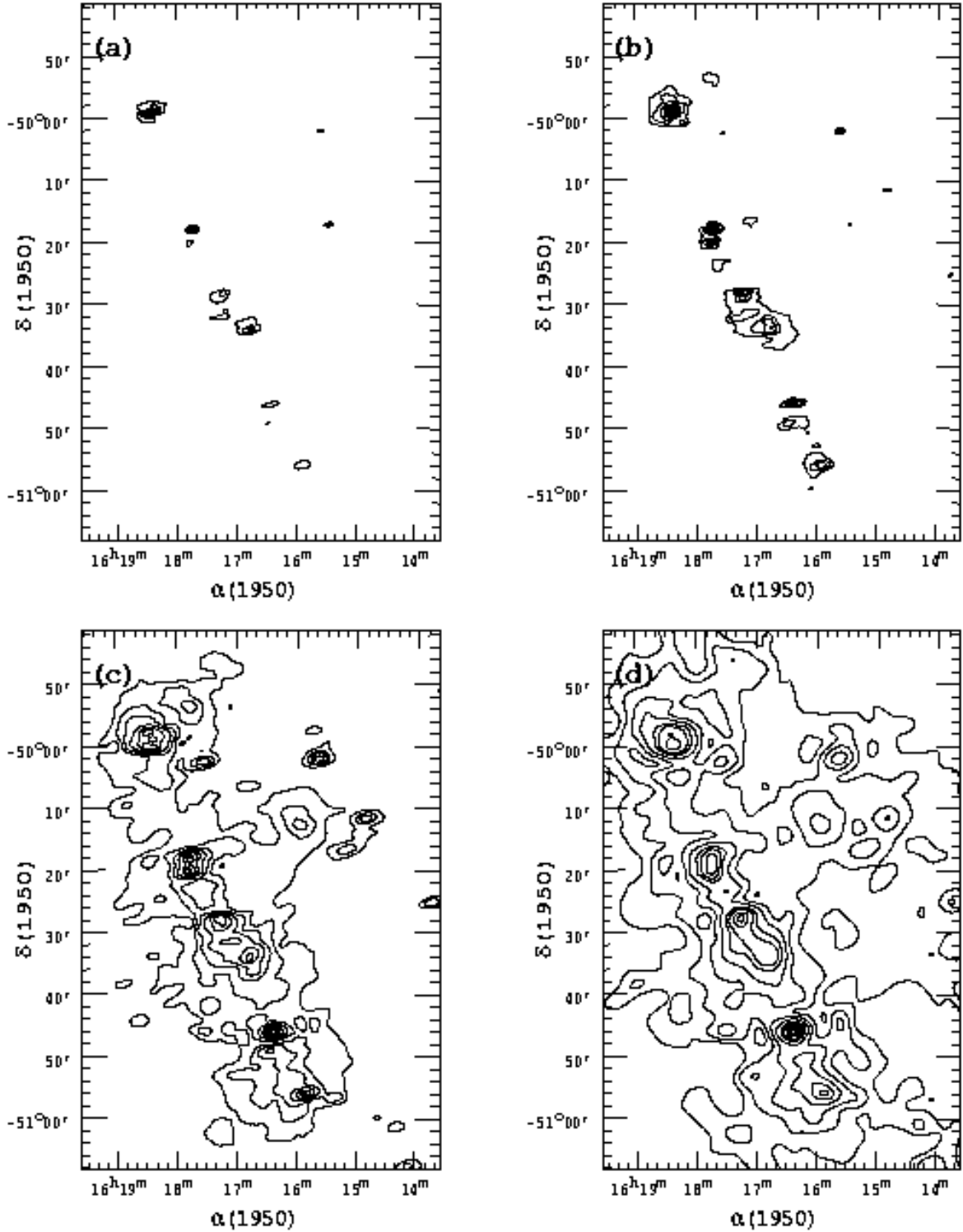,width=16.5cm}}
\caption{ HIRES processed {\it IRAS} maps of RCW 106 region at (a) 12, (b) 25,
 (c) 60 and (d) 100 \micron.
 Contour levels are drawn at 1, 2.5, 5, 10, 20, 30, 50, 70 and 90
percent of peak intensity  4875, 8455, 7947 and 4884 Jy /\sqam\
for 12, 25, 60 and 100 \micron\ respectively.
}
\label{rcwhires}
\end{figure*}

Figs.\ 3a--d show the {\it IRAS} HIRES intensity maps at 12, 25, 60 and 100 
$\mu$m. Once again the lowest contour level shown is 1\% of the peak. 
The {\it IRAS} maps, especially at 60 and 100 $\mu$m are morphologically similar 
to our maps. However, we resolve more sources as compared to the {\it IRAS}.
The FIR maps are also very similar to the radio continuum maps of
Goss \& Shaver (1970) and Shaver \& Goss (1970a).

\subsection{ Maps of temperature and optical depth}
\begin{figure*}
\centerline{\psfig{angle=0,figure=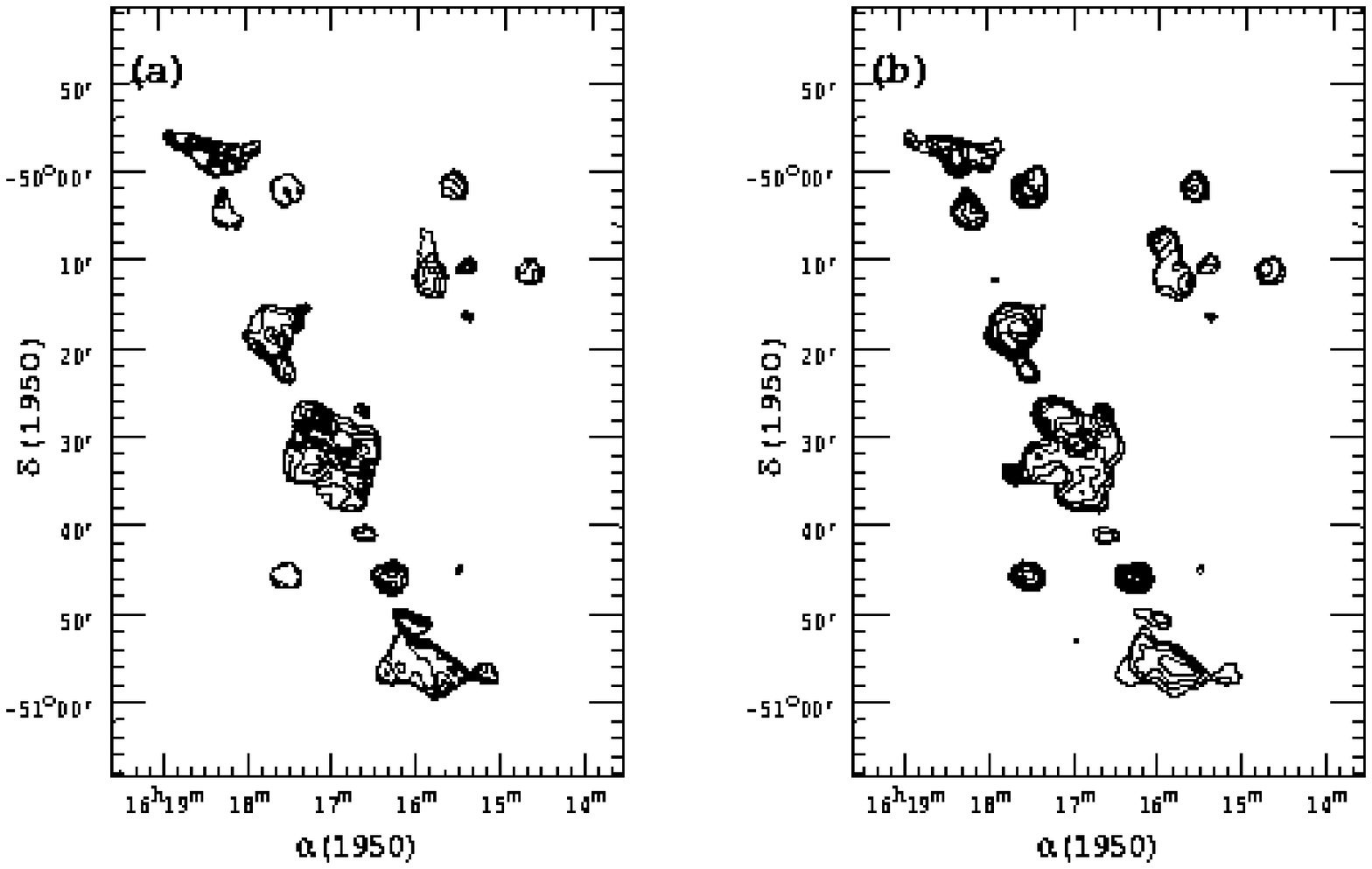,width=18.5cm}}

\caption{ The maps of RCW 106 region for (a) Temperature ($T_{150/210}$)
and (b) optical depth
at 210 \micron.
The dust grain emissivity is assumed to be $\propto \nu^2$.
Contours for temperature are plotted at 20, 24, 28, 32, 36, 40 and 44 K.
Contours for optical depth are plotted at 1, 2.5, 5, 10, 20, 30,
50, 70, and 90 percent of peak optical depth of  0.10 .
\label{Ttaumaps}
}
\end{figure*}

Far-infrared and sub-millimeter observations can be effectively used to
obtain column densities of the region
from the optical depth at these wavelengths  \cite{Hildebrand83}.
Recent studies using {\it IRAS} data at 60 and 100 $\mu$m 
have shown that the 100 $\mu$m optical depth correlates well with other
column density indicators namely CO and visual extinction 
(Langer et al. 1989; Jarrett, Dickman \& Herbst 1989). Since we have
simultaneous observations in two trans-{\it IRAS} wavelengths with almost
identical FOV, we make use of these maps to derive reliable maps of both
the temperature and
optical depth. Our longer wavelength coverage also makes our 
data more sensitive to colder dust (temperature down to about 15 K) as 
compared to the {\it IRAS} coverage. The  intensity maps were smoothed
by taking average within a box of  $0\farcm9 \times 0\farcm9$ size
and these were, in turn, used to obtain maps of temperature and optical
depth. To derive the temperature, we assume an
emissivity dependence \elII\ . These are shown in Fig. \ref{Ttaumaps}.

The temperature map shows a lot of structure and several high temperature
 regions near the cloud boundaries.
These regions  are, most probably, sites of \hii\ blisters where the
dust is heated by  radiation from young stars close to the boundary. The
optical depth map is morphologically similar to the intensity maps and the
 peaks in the two coincide indicating presence of high densities near 
the embedded sources.

 Assuming a homogeneous mixture of  silicate 
($\rho=$ 3.3 g/cm$^{-3}$) and graphite  ($\rho=$ 2.26 g/cm$^{-3}$) 
dust grains with a power law size distribution (index = $-3.5$) 
as  given by Mathis, Rumpl \& Nordsieck (1977) and using the absorption
efficiencies as per Laor \& Draine (1993),  we estimate the total dust
mass of the complex to be equal to 1800 \ms . If one takes the gas to dust
ratio to be 100, then the total mass of the cloud is  $1.8\times 10^5$ \ms .
This value is typical for a GMC of size of 70 pc in the Galactic plane.
However it may be noted that this estimate of cloud mass may be an
underestimate, because due to chopping, low gradient diffuse flux is missed
in our observations (see next subsection).

\subsection{Embedded sources}

Flux calibrated maps (Figs. 2a and 2b) were
processed to extract discrete sources and to obtain their flux densities.  
The HIRES processed maps of the {\it IRAS} survey data  
at 12, 25, 60 and 100 $\mu$m were also used to supplement our observations. 
For all six bands (four {\it IRAS} and two of TIFR)  the following procedure
was adopted to extract sources. For each pixel in the image map, if the  
pixel value was greater than or equal to a specified threshold then the
maximum pixel value within a box of size 5$\times$5 pixels around
this pixel was obtained.  If the pixel value was equal to this maximum, the
position was assumed to be the peak position for a source and its centroid 
position was determined using the IRAF software. A total of 
23 sources were thus found at 210 $\mu$m. More details can be found in Karnik 
(2000). 

The associations of the above sources  with sources in other
maps viz.\ , 150 $\mu$m map of TIFR and the {\it IRAS} maps
at 12, 25, 60 and 100 $\mu$m were searched for, demanding a centroid
separation of less than 1\farcm0 at 150 $\mu$m and 1\farcm5 at other
wavelengths. All 23 sources are detected at 150 $\mu$m. Eighteen of the 23
sources have {\it IRAS} associations with detection in two or more bands. We
have five sources that are detected only in our maps. This is due to
the superior resolution of our maps resulting from the smaller FOV. 
The  flux  densities of sources  detected in each band were obtained
using the numerical aperture photometry 
and an aperture value of 3$\arcmin$ diameter; the
 sky background was evaluated as mode value within an annulus of
 $0 \farcm 5$ width at radius of 4$\arcmin$ and subtracted
from each pixel before obtaining the aperture sum. The  flux densities
 obtained at 150 and 210 $\mu$m bands  were colour corrected
assuming a gray body spectrum with emissivity $\propto \lambda^{-2}$, 
and a temperature corresponding to the ratio of fluxes in the two bands.  
The  {\it IRAS} flux densities were colour corrected
assuming a power law type of flux density distribution between two 
neighbouring bands.
The overall errors on the total flux densities including calibration 
errors are about 15--20\%.
The colour corrected flux densities along with the source positions 
in the 210 $\mu$m map are presented in  Table \ref{rcwsources}. The  SED
of each source was constructed using these tabulated values. The luminosities
of the sources were derived by integrating the SED and assuming a
distance of 3.6 kpc \cite{1979ApJ...232..761L}. Out of band correction
based on the T(150/210)
was also made. For the five sources which were detected only in our
maps the total flux was estimated from the temperature given by the ratio of 
flux densities and assuming \elII . For source S23, {\it IRAS} signals were
saturated at 60 and 100 $\mu$m. We, therefore, estimated these by extrapolation
from the 150 and 210 $\mu$m flux densities. The computed luminosities for
all the sources are also shown in Table \ref{rcwsources}.

\begin{table}
\caption{Positions at 210 $\mu$m and other properties of detected sources}
\label{rcwsources}
\begin{tabular}{crrcrrrrrrrc}
\hline
Source  & R.A.$^{*}$  & DEC$^{*}$ & IRAS PSC  & \multicolumn{6}{c}
{Flux Density in Jy$^{\dagger}$ } & Lum. & IRE$^{\ddagger}$ \\
& (1950) & (1950) & & 12 \mun  &   25 \mun &   60 \mun &  100 \mun& 150 \mun  & 210 \mun & 10$^{3}$ \ls \\
\hline
S1  & 16 14 27.6 & -50 49 40  & -- &   -- &     -- &     -- &     -- &    872 &    326 & 62 & 0.46$^{1}$ \\
S2 & 16 14 43.1 & -50 11 21  & 16148-5011 &   64.6 &    208 &   1822 &   2648 &   1357 &    766 & 52 & -- \\
S3  & 16 15 01.1 & -50 16 02  & - &     -- &   140 &   1229 &   2253 &    648 &    308 & 32  & -- \\
S4 & 16 15 23.3 & -50 16 27  &  16153-5016 &  103 &    140 &   1229 &   2253 &    690 &    342 & 38  & -- \\
S5  & 16 15 31.2 & -50 45 08  & -- &   52.5 &    143 &   1193 &   1432 &  653 &    338 & 32  & -- \\
S6 & 16 15 34.6 & -50 01 46  & 16156-5002 &   102 &    359 &   2930 &   3492 &   2379 &   1275 &  82 & 1.0$^{1}$ \\
S7  & 16 15 42.1 & -50 56 25  & 16158-5055 &   331 &   1732 &   8140 &   8158 &   6373 &   3079 & 249 & 0.55$^{1}$ \\
S8 & 16 15 48.4 & -50 12 29  & 16159-5012 &    54.7 &    257 &   2611 &   3801 &   1910 &    976 & 71 & 0.17$^{1}$ \\
S9 & 16 15 58.6 &  -50 07 53 & -- &  -- &       -- &     -- &     -- &   1332 &    996 & 9.8 & -- \\
S10  & 16 16 00.7 & -50 50 48  & 16159-5049 &   137 &    503 &   3200 &   3713 &   2050 &    817 & 91 & -- \\
S11  & 16 16 18.2 & -50 46 08  & 16164-5046 &   221 &   2033 &  14120 &  16790 &  15380 &   7790 & 411 & 1.04$^{1}$ \\
S12  & 16 16 37.5 & -50 41 17  & -- & -- &       -- &     -- &     -- &   1073 &  512 & 21 & -- \\
S13  & 16 16 43.3 & -50 29 09  & -- &  -- &       -- &     -- &     -- &   2715 & 1275 & 56 & -- \\
S14  & 16 16 58.2 & -50 52 56  & 16170-5053 &  -- &    203 &   1338 &   1485 &    582 &  333 & 31 & -- \\
S15 & 16 17 10.8 & -50 27 48  & 16172-5028 &   468 &   3044 &  12920 &  18355 &  20811 &  10009 & 482 & 0.58$^{1}$ \\
S16  & 16 17 31.5 & -50 45 52  & 16175-5045 &      -- &    101 &   1141 &   1818 &   2372 &   1546 & 37 & -- \\
S17 & 16 17 31.6 & -50 02 20  & 16175-5002 &    40.2 &    177 &   1833 &   3201 &   3198 &   2132 & 62 & 1.50$^{2,3}$ \\
S18 & 16 17 33.3 & -50 22 37  & 16174-5022 &   134 &    600 &   4019 &   5576 &   2180 &   1043 & 115 & -- \\
S19 & 16 17 42.9 & -50 34 03  & -- &  -- &      -- &     -- &     -- &    973 &  1017 & 5.4 & -- \\
S20 & 16 17 42.9 & -50 18 52  & 16177-5018 &   417 &   3164 &  13207 &  17243 &  15572 &   8015 &  460 & 0.72$^{1,2}$ \\
S21 & 16 17 53.3 & -50 12 25  & -- &   -- &     -- &   1088 &   2168 &    572 &  456 & 19 & -- \\
S22 & 16 18 10.4 & -50 04 18  & - & 54.8 &    214 &   1970 &   3940 &   2194 &  1524 & 65 & -- \\
S23 & 16 18 22.9 & -49 58 25  & 16183-4958 &  3372 &  12097 &  12395 &  16785 &  27841 &  11555 & 921 & $>$0.57$^{1,2,3}$ \\
\hline
\end{tabular}
$^{*}$ Units of Right Ascension are hours, minutes and seconds, and the units
of Declination are degrees, minutes and seconds.

$^{\dagger}$ Flux densities are within 3$\arcmin$ diameter.

$^{\ddagger}$ References of radio data for calculating IRE are -- 1)
Shaver \& Goss (1970b); 2) Retallack  \& Goss (1980); 3) Forster et al.
(1987).

\end{table}

 We  estimate the diffuse emission in the complex by subtracting the 
sum of flux densities of the extracted sources from the total flux 
density of the map down to 1\% contour level. 
We find the diffuse fraction to be 81\%, 84\%, 19\% and
27\% at 60, 100, 150 and 210  $\mu$m respectively. In contrast to the {\it IRAS},
our chopped observations are less sensitive to low gradient diffuse flux.
The fraction of diffuse flux obtained using chopped observations of RCW 106
complex may be compared with values obtained for other regions using similar
chopped observations: $\leq$ 50\% for Carina complex \cite{Ghosh88}, 35\%
for W 31 \cite{Ghosh89b} and 35\% for {\it IRAS} 09002$-$4732 region \cite{Ghosh00} .

\subsection{Comparison with radio observations}

We have looked for association of the FIR sources with radio continuum
sources detected by others, demanding that positions match within
$2\farcm 0$. We find 10 associations. Taking the FIR luminosity listed in
Table 1 as that due to a single ZAMS star, we can estimate the expected rate
of emission of Lyman continuum photons. We can also compute the same from
the observed radio continuum flux density using the formula of Mezger
(1978). We have used the 5000
MHz flux density given by Shaver \& Goss (1970b) (at an angular
resolution of 4$\arcmin$) wherever available since at
this high frequency the radio emission is less affected by optical depth
effects. For source S17 we have used the mean of the 1415 MHz flux density
of Retallack \& Goss (1980) and 1420 MHz flux density of Forster et al.
(1987) with an angular resolution $\sim$~1$\arcmin$ . We have
not used very high resolution ($\sim$~1$\arcsec$) interferometric
radio observations of Walsh et al. (1998) as they only give
peak flux densities. In Table
1 we present IRE, the ratio of Lyman continuum luminosity derived from
the FIR and radio observations.
If the radio emission is optically thin and the source is
powered by a single ZAMS star, IRE = 1. If however, the energizing source is 
multiple or if some fraction of Lyman continuum luminosity is absorbed by
dust and not available for ionizing the gas, we expect IRE $>$ 1. This
can also result if a fraction of the FIR luminosity is due to accretion
(Garay \& Lizano 1999) since accretion is unlikely to produce energetic
Lyman continuum photons. From Table 1 we find that only one source viz.,
S17 has IRE significantly higher than unity and even for this, it is
consistent with unity if 50 \% of Lyman continuum photons are unavailable
for ionization because of dust absorption (Smith, Biermann \& Mezger 1978).
For all other sources the ratio is $\leq$ 1; this is most probably due
to the fact that the resolution of radio observations used is generally
much poorer than that of our FIR observations.

It is generally believed that massive stars are formed in clusters.
We may, therefore, have additional unresolved stars within our effective
resolution of 1$\farcm2$. However, in the light of IRE discussed above, we
cannot make definite statement about the presence of multiple sources. We
can only conclude that another massive star with a
significant fraction of observed luminosity is unlikely to be present.
Similar  statement can also be made for luminosity from accretion.    

\section{Stellar Luminosity Distribution and Initial Mass Function}

\begin{figure}
\leavevmode
\hspace*{1.5in}
\epsfxsize=250.0pt
\epsfysize=250.0pt
\epsfbox{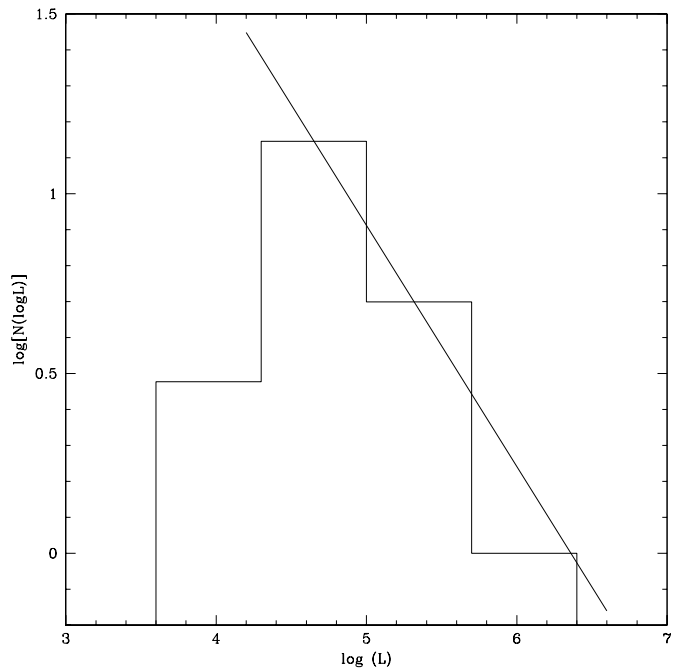}
\caption{
Luminosity distribution for sources in RCW 106 region.
The line shows the fit obtained using
maximum likelihood method assuming power law type of number distribution. 
\label{lumfig}}
\end{figure}

For deeply embedded objects like the RCW 106, the FIR luminosity is 
almost the bolometric luminosity of the energizing star(s). Making use of the 23 
objects detected by us, we construct the stellar luminosity distribution of 
massive stars in the RCW 106 complex. This is shown in Fig. \ref{lumfig}.
 From an
inspection of the figure we see that our sampling is nearly complete for
L $>$ 20,000 \ls . Assuming the luminosity distribution to be a power law
given by N[log (L/\ls )] $\propto$ $\alpha$ log (L/\ls ) and using the more
robust (as compared
to binning) maximum likelihood method, we find $\alpha$ = $-0.67\pm0.22$
for L $>$ 20,000 \ls . It may be noted that due to factors discussed in
the previous section, the observed distribution could be flatter than
the intrinsic one. Our value of slope may be
compared with the value of $-0.56$ derived by Rengarajan (1984) from the
M17SW data of Jaffe, Stier \& Fazio (1982).

%
%

The initial mass function  (IMF) of stars is usually defined as
the birth rate of  stars formed per unit logarithmic mass interval per unit
time. The shape of the IMF ($\xi (log~m_*)$) is generally characterized as 
a power law with
a spectral index ($\beta$). The IMF at birth is normally derived from 
the observed present day mass function (PDMF) and applying corrections
 to it. For example, the well known Salpeter IMF \cite{Salpeter55}   
and Miller Scalo IMF \cite{MillerScalo79} are derived from the
 population of field stars after correcting for life times, scale 
heights etc. These estimates are, therefore, averaged over a large 
volume and over the life of the Galaxy. In a recent review  Scalo (1998) 
summarises the more accurate spectroscopic observations of massive stars in 
clusters and associations (age up to a few million years) and concludes that
there is a large uncertainty in the value of $\beta$ and that it ranges
between $-1$ and $-$2. 
An interesting question is what is the shape of IMF at a very early 
stage when stars are still deeply embedded in molecular clouds? The 
large number of high mass stars observed in the RCW complex and a 
young age ($<$ 100,000 years) give us an opportunity to study the IMF at
birth. For this purpose, the FIR luminosities were
 used to obtain masses using the mass--luminosity relation
for a ZAMS star with solar metallicity, using the analytical
relation given by Tout et al. (1996). Again using the maximum likelihood
method we find $\beta$ = $-1.73\pm0.5$ for mass $\geq$ 15 \ms . 
As is the case for the luminosity distribution, the actual IMF would
be steeper. This slope is in the same region as the values determined for
massive stars and
for stars of intermediate mass from photometric observations of stellar
clusters (Scalo 1998). Ghosh et al. (2000) obtained  $\beta$ = $-1.25$ in
the mass range of 4--16 \ms\ in the complex around {\it IRAS} 09002$-$4736.
Okumura et al. (2000) using near infrared imaging observations of W51 find
that the slope above 10 \ms\ is consistent with $-1.8$ and perhaps flattens
above 30 \ms .

\section{Radiation Transfer Modelling}

\subsection{Modelling Procedure}

\begin{table}
\caption{Parameters$^*$ for the models shown in Figures \ref{sedplot1} and
\ref{sedplot2}}
\label{bestmodpar}
\begin{tabular}{llllllllllllllllll}
\hline
 & \multicolumn{5}{c}{Best fit} & \multicolumn{6}{c}{Second best fit} &
 \multicolumn{6}{c}{Poorest fit} \\
Source & R  & R$_{c}$   & $\gamma$ & $\tau_{100}$ &  f$_{sil}$
& R  & R$_{c}$ & $\gamma$ & $\tau_{100}$ &  f$_{sil}$ & $\chi^{2\dagger}$ 
& R  & R$_{c}$ & $\gamma$ & $\tau_{100}$ &  f$_{sil}$  & $\chi^{2\dagger}$ \\
 & (pc) & (pc) & & & & (pc) & (pc) & & & & & (pc) & (pc) \\
\hline
S2 & 0.64 & 0.017 & 1 & 0.19 &  0.13 & 0.68 & 0.009 & 0 & 0.06 &  0.66 & 1.0
 & 0.60 & 0.064 & 2 & 0.17 &  0.10 & 4.7\\
S4 & 0.26 & 0.016 & 1 & 0.21 &  0.00   & 0.38 & 0.008 & 0 & 0.07 &  0.23 & 1.3 
 & 0.30 & 0.034 & 2 & 0.20 &  0.00 & 1.8 \\ 
S5 & 0.37 & 0.006 & 0 & 0.07 &  0.39  & 0.39 & 0.018 & 1 & 0.16 &  0.16 & 1.4
 & 0.38 & 0.040 & 2 & 0.16 &  0.10 & 2.9\\ 
S6 & 0.81 & 0.009 & 1 & 0.25 &  0.07  & 1.04 & 0.013 & 0 & 0.05 &  0.89  & 6.1
 & 0.90 & 0.087 & 2 & 0.17 &  0.06 & 40.0  \\ 
S7 & 2.14 & 0.038 & 1 & 0.10 &  0.33  & 1.22 & 0.028 & 0 & 0.06 &  0.56 & 5.4
 & 1.10 & 0.153 & 2 & 0.12 &  0.05 & 6.8\\ 
S8 & 0.56 & 0.021 & 1 & 0.22 &  0.12   & 0.61 & 0.009 & 0 & 0.09 &  0.55  & 1.7
 & 0.44 & 0.070 & 2 & 0.20 &  0.09 & 5.7 \\ 
S10 & 0.58 & 0.021 & 0 & 0.07 &  0.45  & 0.87 & 0.026 & 1 & 0.11 &  0.28 & 1.5
 & 0.56 & 0.009 & 2 & 0.13 &  0.14 & 3.7\\  
S11 & 2.62 & 0.054 & 1 & 0.18 &  0.10  & 2.08 & 0.182 & 0 & 0.09 &  0.40 & 1.5
 & 2.77 & 0.168 & 2 & 0.18 &  0.07 & 11.0\\ 
S15 & 4.36 & 0.051 & 1 & 0.15 &  0.20  & 2.42 & 0.017 & 0 & 0.08 &  0.47 & 7.6
 & 2.13 & 0.219 & 2 & 0.13 &  0.06 & 11.0\\ 
S17 & 1.24 & 0.016 & 0 & 0.11 &  0.48   & 1.06 & 0.040 & 1 & 0.25 &  0.08  & 4.6
 & 0.72 & 0.146 & 2 & 0.22 &  0.06 & 22.6 \\ 
S18 & 0.69 & 0.022 & 1 & 0.15 &  0.18 & 0.69 & 0.016 & 0 & 0.06 &  0.75 & 1.2
 & 0.50 & 0.093 & 2 & 0.14 &  0.95 & 1.8\\ 
S20 & 4.32 & 0.063 & 1 & 0.11 &  0.36   & 1.93 & 0.010 & 0 & 0.07 &  0.51 & 19.0
 & 1.13 & 0.212 & 2 & 0.16 &  0.10 & 24.3 \\ 
S22 & 0.78 & 0.024 & 1 & 0.27 &  0.06  & 0.84 & 0.014 & 0 & 0.10 &  0.47 & 1.3
 & 1.76 & 0.083 & 2 & 0.22 &  0.03 & 12.2\\ 
\hline                          
\end{tabular}                  

$^*$For each source three sets of parameters are given. These are for the
best fit models for each of the three types of density distributions. The
order among the three sets is according to the fit. Thus the first set
corresponds to the best fit out of the three, the last corresponds to the
poorest fit. The parameters are -- R, the radius of the cloud, R$_{c}$ the
radius of the central dust free cavity, $\gamma$, the power law index of the
radial density distribution [$\rho(r)\propto r^{-\gamma}$], $\tau_{100}$,
the radial optical depth at 100 $\mu$m and f$_{sil}$, the fractional
abundance of silicate.\\

$^{\dagger} \chi^{2}$ values are given relative to
the best fit (i.e. $\chi^{2}_{\rm Second~ best~ fit}/\chi^{2}_{\rm Best~ fit}$
and $\chi^{2}_{\rm Poorest~ fit}/\chi^{2}_{\rm Best~ fit}$ ).

\end{table}

The emergent SED of an embedded source depends on the luminosity of
 the source and  the physical properties of its dusty envelope. By 
fitting the observed SED to radiation transfer calculation, one can 
obtain useful information on the geometric details and nature of dust.
 With this in mind we have chosen, from Table 1, thirteen sources that
 have measured flux densities in all six bands, viz.\ , the two TIFR 
and four {\it IRAS} bands. The source S23 was excluded as the {\it IRAS} 
 signals at 60 and 100 $\mu$m are saturated.
The radiation transfer calculation was carried out through a purely dust
component in a spherical geometry using the code CSDUST3 \cite{csdust3}.
The  central source was assumed to be a single star with a luminosity 
equal to the value derived from observations. 
A mixture of graphite and astronomical silicate with properties as given by 
Laor \& Draine (1993) was assumed.  Size distribution [$n(a)$]  was assumed 
to be a power law [$n(a) \propto a^{-3.5}$] as per Mathis et~al.
(1977). The free parameters of the model were  -- R, the size of the
envelope , R$_{c}$, the size of the central dust free cavity , $\tau_{100}$,
the total (due to both type of grains) radial optical depth at 100 \m ,
$f_{sil}$ , the fractional abundance of silicate   
and $\gamma$, the power law index  of the radial density 
distribution within the envelope [$\rho(r)\propto r^{-\gamma}$].  
Only three discrete values of 0, 1 and 2 were explored for $\gamma$.
A $\chi ^{2}$ minimization scheme was developed  for
fitting the computed SED of the
model to the  observed SED of  the source and the fitting was done 
for each value of $\gamma$.  
 
The sensitivity of the $\chi^2$ minimization scheme
in terms of extracting the parameters (mainly for index $\gamma$)
accurately from the input 
observed SED of the source was tested using simulation. 
Synthetic input was created by calculating the flux densities at the six 
wavelengths used as well as in the NIR (5 \m )  and sub-mm  (800 \m ) 
with added random noises. From
 the results of the simulation we conclude that i) the best fit  always 
corresponds to the case when the $\gamma$ value is the same for the 
input spectrum and  the modelled spectrum; presence of NIR point 
improves the discrimination especially for $\gamma$ = 1 or 2; ii) the 
radius of the central dust free cavity and optical depth at 100 $\mu$m
are the most sensitive parameters of the modelling procedure, the variation
between the input parameters and the extracted parameters for both is about 
20\%, while for the outer radius it is about 30\%; iii) the dust 
composition is a parameter with poor extraction accuracy since 
the absorption efficiency beyond about 50 $\mu$m is the same for graphite 
and silicate and the main constraints come from data points at 25 $\mu$m 
or lower.

\subsection{ Results of modelling RCW 106 sources}

\begin{figure*}
\epsfysize=400.0pt
\epsfbox{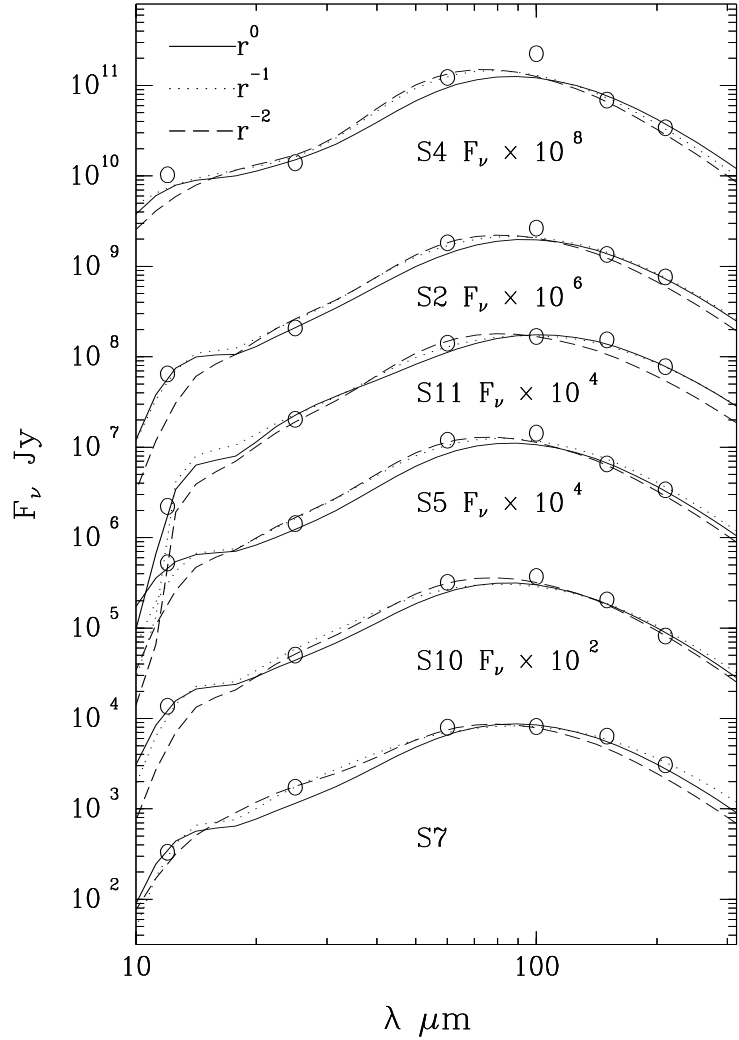}

\caption{Spectral energy distribution for the six sources S7, S10, S5,
S11, S2 and S4
in RCW 106 complex region as given in Table \ref{rcwsources}.
The lines shown are fits from radiation transfer models. Different lines
depict fits for three different power law dependence of 
radial density. The parameters for these fits are given in Table
\ref{bestmodpar}. Note that for the sake of clarity
flux densities have been multiplied by a 
different constant factor for five of the sources.
}
\label{sedplot1} 
\end{figure*}

\begin{figure*}
\epsfysize=400.0pt
\epsfbox{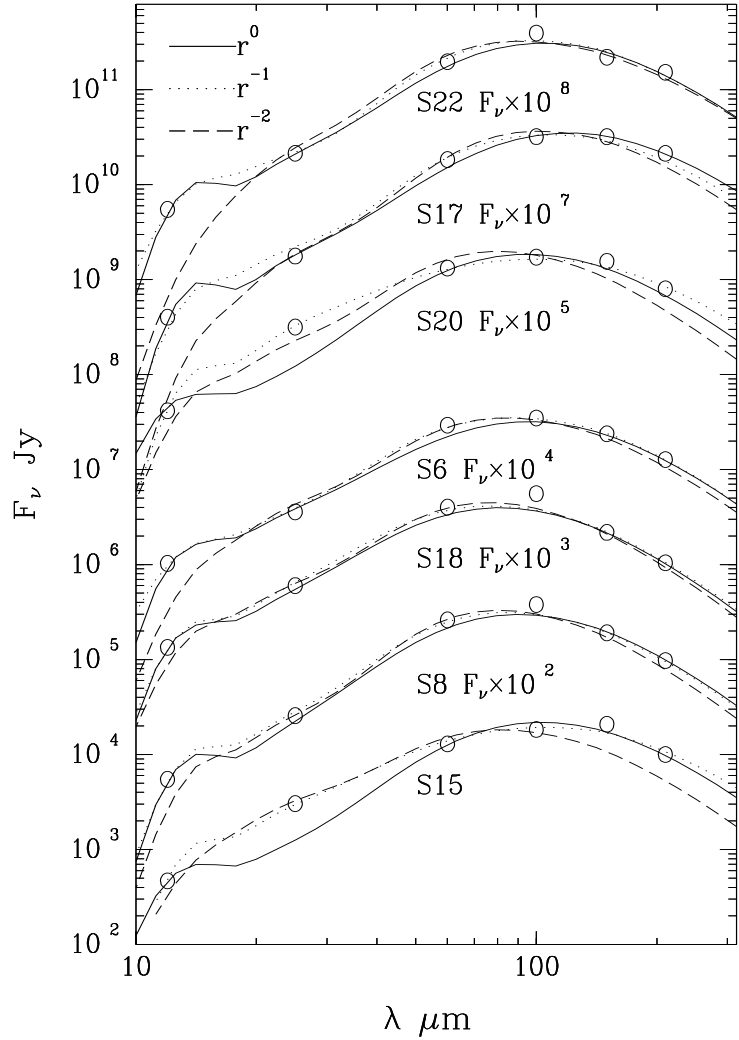}

\caption{
SED plots similar to those in Fig. \ref{sedplot1}, but for sources 
S15, S8, S18, S6, S20, S17 and S22.
}
\label{sedplot2}
\end{figure*}

Figures \ref{sedplot1} and \ref{sedplot2} show the observed data points
and the best fit SED for each value of $\gamma$ for the 13 selected sources.
The parameters of these models, along with relative $\chi ^{2}$  values, are
presented in Table \ref{bestmodpar}.

\begin{table}
\caption{Density distribution near embedded high mass stars}
\label{raddist}
\begin{center}
\begin{tabular}{lll}
\hline
 $Resolution$ & $\gamma$   &  Reference \\
\hline
 $1 \farcm 2$ & 0 or 1 & (FIR) Present work, Ghosh et al. (2000), \\
\smallskip
 & & Mookerjea et al. (1999,2000) \\
\smallskip
 $ 0\farcm5$ & 0 or 1 & (FIR) \cite{Campbell95}\\
\smallskip
 $0\farcm15$ &    0        & (MIR) \cite{Faison98}\\
\smallskip
 $0\farcm15$/$0\farcm25$ &    1.5  & (Sub-mm) \cite{Hatchell00}\\
\smallskip
 $0\farcm5$  &   2--1        & (Sub-mm) \cite{Hunter97}\\
\hline
\end{tabular}
\end{center}
\end{table}

From Figs. \ref{sedplot1}, Fig. \ref{sedplot2}  and Table 2 it is seen
 that for 9 out of the 13 sources  $r^{-1}$ distribution fits the SED 
best while 
for the rest, uniform distribution is the best fit. 
It may be mentioned that there is not much difference between the
the final $\chi^2$ obtained for the best fits for the
two density distributions. The $r^{-2}$ distribution
shows the poorest fit  for all sources.
Table \ref{raddist} lists the radial
distribution index obtained in recent times by different authors using
  radiation transfer modelling
of envelopes of high mass embedded sources. This compilation suggests
that in many cases the density distribution differs from
$\rho(r) \propto r^{-2}$ or $r^{-1.5}$ that is
 expected for the envelopes of young protostars \cite{ShuAdamsLizano}.
Is there really a lack of sources with $r^{-2}$ density distribution as
revealed by FIR and MIR observations?
This can be checked only through higher
angular resolution observations at FIR wavelengths and
modelling actual radial intensity profiles.
However, there is no prospect of large improvement 
of angular resolution in the FIR in the near future.
If the density distribution is indeed different from 
that expected for free fall models (i.e. $\propto r^{-2}$),
what are the possible reasons for this?
Do high velocity outflows arising from young high
mass stars and  radiation pressure on dust grains due to intense 
UV radiation from the young high mass star help in  modifying
the initial free fall density distribution $\propto r^{-2}$ ? 
Jijina \& Adams (1996), accounting for effect of  radiation pressure,
predict that close to the star, density distribution should
be $\propto r^{-0.5}$ and at larger distances $\propto r^{-1.5}$.
One should note that these solutions are at much smaller scales 
compared to those sampled by the large beams  of most of the FIR observations. 

\subsection{Luminosity and Envelope Mass} 

\begin{figure}
\centerline{\psfig{angle=0,figure=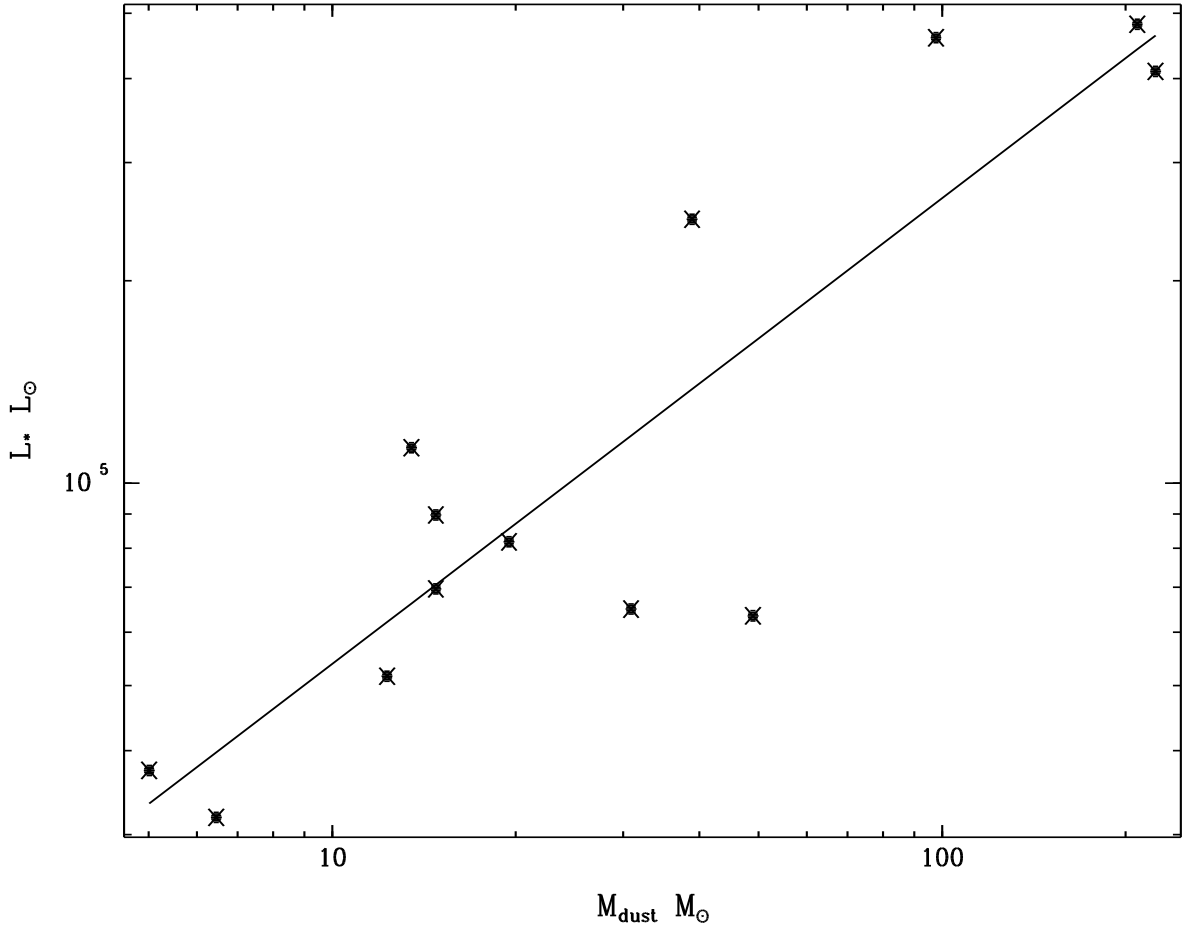,width=16.5cm}}
\caption{
The FIR luminosity of the source 
plotted against the mass of the dust envelope obtained from the best fit 
radiation transfer 
model. The solid line represents the power law regression fit to the points.
\label{lumin-mass}
}
\end{figure}
  The mass computed from the best fit model parameters represents
the mass of the star forming envelope if one assumes a constant gas to 
dust ratio. 
In Fig. \ref{lumin-mass} the luminosity of the source is plotted
against the computed dust mass of the envelope.
A correlation is seen between the two.
The probability of
finding no correlation  is as low as 0.0086. 
The  power law  index after regression analysis is found to be $0.7\pm0.11$.
This indicates
that the more massive molecular cores  produce more massive stars.

\section{Summary}

The star forming region associated with RCW 106 has been mapped
simultaneously at 150 and 210 \micron. The HIRES processed {\it IRAS} survey
data are used as supplement to our data. The temperature and the FIR
optical depth maps have been generated for the region. The total dust
mass estimated from the optical depth map is about 1800 \ms . The local
maxima in the 210 \micron\ map have been searched for to find the embedded
sources. The associations of these sources are found in the four {\it IRAS} maps
and the 150 \micron\ map. 
The diffuse fraction in these maps is found to be 81\% and 84\% in the
{\it IRAS} 60 and 100 \micron\ bands while it is 19\% and 
27\% at 150 and 210 \micron\ from our chopped observations.
The SEDs for the sources are constructed and FIR luminosities estimated.
Luminosity distribution of these sources has been obtained. 
These FIR luminosities are used, along with mass-luminosity 
relation for ZAMS stars, to obtain IMF of the region. The estimated
power law index of IMF is $-1.73$ for mass $\geq$ 15 \ms\, similar to
that of other star forming regions and OB associations albeit  flatter
than the  IMF for field stars.
 The SEDs of the sources detected in all
the bands are modelled using radiation transfer calculations
assuming a dusty envelope surrounding the young ZAMS star
with luminosity equal to that of the FIR luminosity of the source.
The size of the envelope, inner dust free core radius, line of sight
optical depth due to dust at 100 $\mu$m and composition of the 
dust were varied to obtain the best fit to the observed SED. Three
types of radial density distributions --
[$\rho(r) \propto r^{-n}$ where $\gamma$=0, 1 and 2] were tried for. 
The best fit
results show that $r^{-2}$ can be ruled out against uniform
and $r^{-1}$ distributions. A correlation was found between the
the source luminosity and the mass of dusty envelopes calculated 
using the best fit model parameters.

\centerline{\bf Acknowledgements}

We thank 
S.L. D'Costa, M.V. Naik, S.V. Gollapudi, D.M. Patkar, 
M.B. Naik and G.S. Meshram for their support for the
experiment. The members of TIFR Balloon Facility (Balloon group and
Control \& Instrumentation group), Hyderabad, are thanked for their roles
in conducting the balloon flights. IPAC is thanked for providing HIRES
processed {\it IRAS} data. TNR thanks JSPS and Dr. H. Shibai for the Invitation
Fellowship at the Physics Department, Nagoya University where part of this
work was done.

\end{document}